\documentclass[10pt,conference]{IEEEtran}
\usepackage{cite}
\usepackage[tight,footnotesize]{subfigure}
\usepackage{color}
\usepackage[normalem]{ulem}
\usepackage{algorithm}
\usepackage{multicol}
\usepackage[numbers,sort&compress]{natbib}
\usepackage{algpseudocode}
\usepackage{url}
\usepackage{breakurl}

\ifCLASSINFOpdf
   \usepackage[pdftex]{graphicx}
\else
\fi
\usepackage[cmex10]{amsmath}
\usepackage{amssymb}

\usepackage[singlelinecheck=off]{caption}

\usepackage{url}

\begin{document}

\title{Skin-Health Monitoring system using a Wireless Body Area Network}

 
\author{\IEEEauthorblockN{Suman Kumar\IEEEauthorrefmark{1},
 Kazi Amanul Islam Siddiqui\IEEEauthorrefmark{1}, Mukesh Kumar\IEEEauthorrefmark{2}}
    \IEEEauthorblockA{\IEEEauthorrefmark{1}Department of Computer Science, Troy University, Troy, AL, USA 
 }
\IEEEauthorblockA{\IEEEauthorrefmark{2}Independent Researcher\\
Emails: \{skumar, ksiddiqui\}@troy.edu, Mukesh\_kumar2000@yahoo.co.uk}
}
%

\maketitle

%

\begin{abstract}
A new class of sensing paradigm known as lab-on-skin where stretchable and flexible smart sensor devices are integrated into the skin, provides direct monitoring and diagnostic interfaces to the body. Distributed lab-on-skin wireless sensors have the ability to provide continuous long term assessment of the skin health. This paper proposes a distributed skin health monitoring system using a wireless body area network. The system is responsive to the dynamic changes in the skin health, and remotely reports on the same. The proposed algorithm detects the abnormal skin and creates an energy efficient data aggregation tree covering the affected area while putting the unnecessary sensors to sleep mode. The algorithm responds to the changing conditions of the skin by dynamically adapting the size and shape of the monitoring trees to that of the abnormal skin areas thus providing a comprehensive monitoring. Simulation results demonstrate the application and utility of the proposed algorithm for changing wound shapes and sizes.

\end{abstract}


%
\IEEEpeerreviewmaketitle

\section{Introduction}
\label{intro}
Smart wearables such as smart watches, fitness trackers, and various specialty smart physiology monitors for continuous health measurements are increasingly becoming widespread, thanks to the advances in miniaturization technology, and lower cost. Subsequently, body wearable sensors have been enabling ubiquitous continuous remote health monitoring. Continuous and comprehensive health monitoring is becoming a reality because sensing and implant technologies have been seeing leaps and bounds of development, especially in the measurement of vital signs and body activities and in holding power and fast recharging capacity~\cite{doi:10.1002/adfm.201504755}. One of the most notable advances is the development of stretchable and flexible sensor devices which can be integrated into skin~\cite{doi:10.1021/acsnano.7b04898}. Being interfaced with the soft skin tissues, the sensors provide direct interfaces to the body, potentially replacing the conventional monitoring and diagnostics methods with a new technology paradigm, coined as Lab-on-Skin. Various medical application domains such as dermatology, cardiology, blood diagnostics, etc. are bound to benefit from it enabling better and cheaper healthcare for everyone. This paper proposes a skin health monitoring system using lab-on-skin technology.

OECD reports that health care cost is rising at a faster rate than the GDPs in developed nations and currently, it is on track to become unsustainable\footnote{https://www.oecd.org/health/}. Early detection of health conditions is extremely important in saving lives and improving health by stopping and/or slowing down the spread of the illness~\cite{Hawkesl408, SurvivalAsympDistMetaCan}, resulting in reduction of healthcare costs\footnote{www.cdc.gov/chronicdisease/about/costs/index.htm}. Early detection leads to timely diagnosis, and both, detection and timely diagnosis require continuous health monitoring. Ubiquitous monitoring can be achieved by integrating low power wireless sensor nodes equipped with computing and communication abilities forming a Body Area Network (BAN). A network of sensors (i.e. BAN) collects physiological data and delivers it to  the base stations, furthermore, base stations (i.e. sink) can distribute the data to remote locations for further processing and analysis. Thus BAN facilitates long term continuous monitoring without affecting the daily activities of a person. Different health conditions demand BAN systems be designed differently to meet its general and special requirements(See~\cite{doi:10.1080/03091902.2020.1729882}). For example, long network life time is a general requirement; on the other hand, health monitoring of a pregnant woman imposes special requirements where BAN must adapt to the health changes which come about with different phases of pregnancy~\cite{8651768}. Likewise, wireless BAN design for ubiquitous skin-health monitoring is not only uniquely challenging but also it presents some great opportunities as described in the next section.  

The rest of the paper is organized as follows: Motivation and contribution of this work are presented in Section~\ref{motiv}. Section \ref{related} presents related work and a brief overview of Lab-on-skin technology. The proposed system is described in Section~\ref{Segalg}. Section~\ref{Evaluation} demonstrates and evaluates this work. Finally, the paper is concluded in~\ref{conclusion} with possible future work.
\begin{figure*}[h] 
  \centering
  \subfigure[Healthy Skin]{\label{HealthySkin}
   \includegraphics[width=0.22\textwidth, height=0.10\textheight]{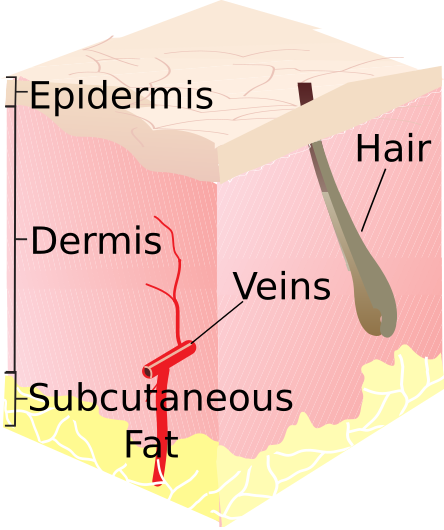}\hfil} 
	\subfigure[Melanoma: Phase I]{\label{Melanomal}
   \includegraphics[width=0.22\textwidth, height=0.10\textheight]{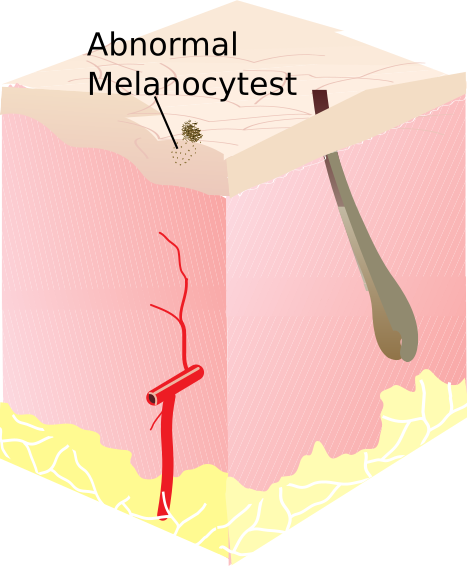}} 
  \subfigure[Melanoma: Phase II \& III]{\label{MelanomaIIIII}
   \includegraphics[width=0.22\textwidth, height=0.10\textheight]{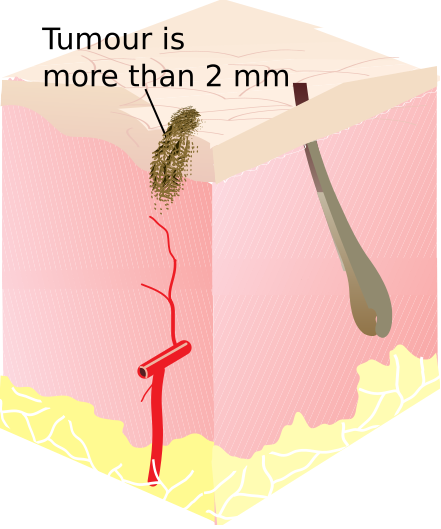}} 
	\subfigure[Melanoma: Phase IV]{\label{MelanomaIV}
   \includegraphics[width=0.22\textwidth, height=0.10\textheight]{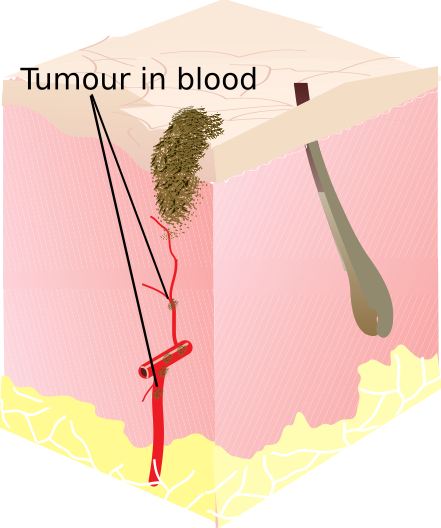}} \\
  \subfigure[Wound: Hemostatis]{\label{WoundI}
   \includegraphics[width=0.22\textwidth, height=0.10\textheight]{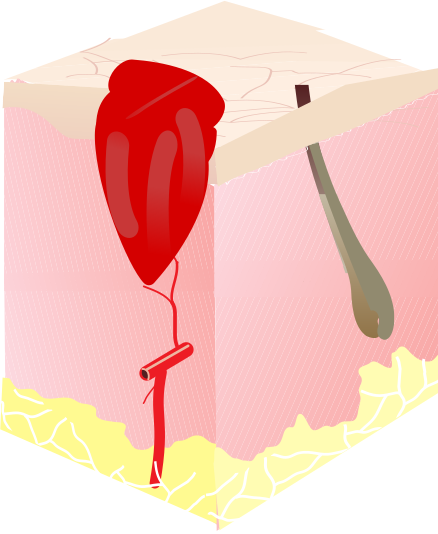}\hfil} 
	\subfigure[Wound: Inflammation]{\label{WoundII}
   \includegraphics[width=0.22\textwidth, height=0.10\textheight]{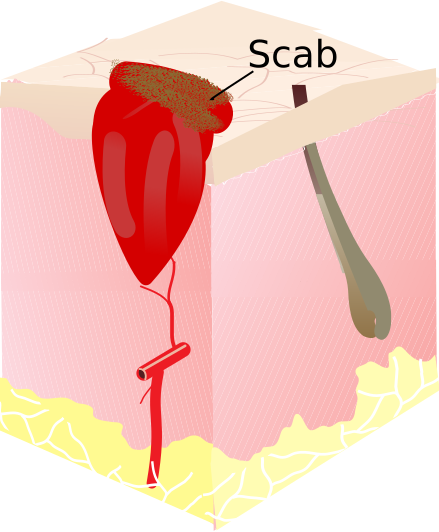}} 
	\subfigure[Wound: Proliferation]{\label{WoundIII}
   \includegraphics[width=0.22\textwidth, height=0.10\textheight]{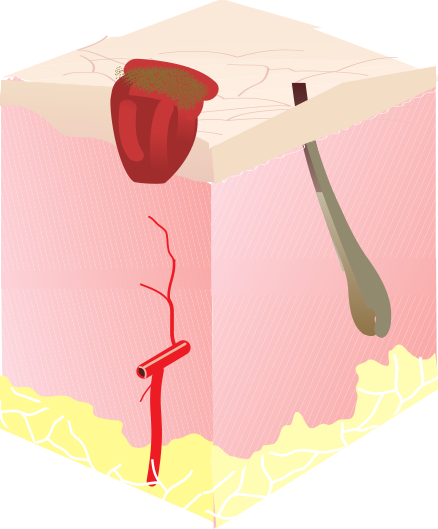}\hfil} 
	\subfigure[Wound: Remodeling]{\label{WoundIV}
   \includegraphics[width=0.22\textwidth, height=0.10\textheight]{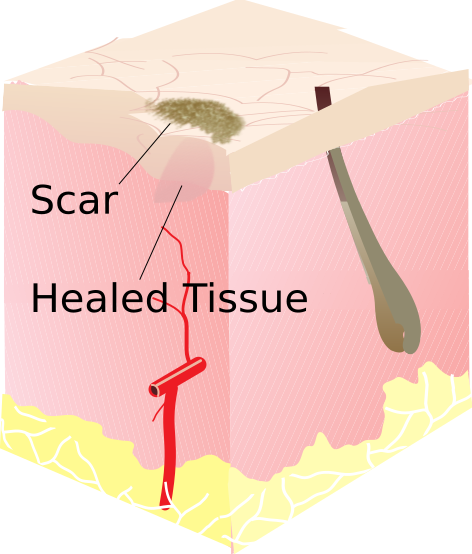}} 
    \caption{The spread and shrinkage of two health conditions, Melanoma, from (b) to (d) and wound caused by a mechanical injury, from (e) to (h) are shown. Melanoma spreads whereas wound shrinks}
  \label{WoundMelanoma}
\end{figure*}

\section{Motivation and Contribution}
\label{motiv}
There are many skin conditions ranging from a simple rash to wound. Skin is the largest organ which is exposed to the external environment; therefore, it is at the highest risk of getting infected, injured, and developing cancer as compared to any other organ. Moreover, CDC estimates that there are more than 13 million workers risk skin exposure to harmful chemical agents\footnote{www.cdc.gov/niosh/topi-cs/skin/default.html} which can result into various serious health conditions. The issue of skin health monitoring is extremely important since skin diseases are widespread which affect mental health conditions leading to addictions which lead to more health problems and the cycle continues. Sadly, the person suffering from skin problems often faces social stigma and alienation.  

Various skin related health conditions exhibit unique characteristics requiring special design considerations while designing monitoring systems. To motivate the need for a unique approach for health monitoring of skin, two skin conditions- Melanoma, a type of a skin cancer and a wound, a type of circumscribed skin injury  are examined as shown in Figure~\ref{WoundMelanoma}. Healthy skin is shown in Figure~\ref{HealthySkin}. Figures~\ref{Melanomal} -~\ref{MelanomaIV} show various stages of Melanoma whereas healing phases of a wound are shown in Figures~\ref{WoundI} - ~\ref{WoundIV}. In the first phase, Melanoma\footnote{www.cancer.org/cancer/melanoma-skin-cancer/} may start with a mole of changing size (abnormal melanocytest), shape and color (See Figure~\ref{Melanomal}). In the second phase, size and depth are increased as shown in Figure~\ref{MelanomaIIIII}. Furthermore, Melanoma progresses to stage 4 if the tumor is spread to the body organs, blood stream and lymph nodes (see Figure~\ref{MelanomaIV}). The healing process of Wound consists of four phases. In the hemostatis phase (Figure~\ref{WoundI}), blood clot starts to form then in second phase, known as inflammatory, scab formation takes place and antibodies, and growth factors fill up the wound area as shown in Figure~\ref{WoundII}. Building of lost or damaged tissues take place in phase III also known as proliferation (Figure~\ref{WoundIII}) and finally, in the final phase, the wound is almost healed and a scar is left.  

Both, Melanoma and Wound can increase and decrease in sizes. Important factors like oxygen level, ph level, sweat, temperature etc. need to be measured continuously to observe status of skin conditions. Since affected skin area can spread or shrink in sizes, a health monitoring system must continuously respond to and report on these changes. Moreover, wireless body area network with body sensors limited in battery power while operating in uncertain conditions can be made to last long if it is designed to adapt to these changes. Such systems not only provide real time monitoring and feedback mechanisms but also improve our understanding of characteristics and factors impacting the skin health conditions by analyzing the collected long term spatio-temporal data~\cite{PACHECO2020103545}. It can lead to better models, helping us design appropriate treatment methodologies as well as the development of better medicines for prevention, detection and management of skin ailments. More importantly, the system enriches Teledermatology which is becoming a necessity and a preferred choice for many in the era of social distancing and isolation due to Covid-19~\cite{doi:10.1111/ced.14280}.

We present, to the best of our knowledge, the first distributed skin health monitoring system which dynamically adapts to the changing affected skin areas while reporting on the same. The contributions of this work are as follows.
\begin{itemize}
	\item  A novel distributed algorithm for continuous skin health monitoring is presented. An aggregation tree topology is constructed to collect the data from the sensors in the affected skin area. The algorithm selects the highest energy node as a root of the aggregation tree and keep only those sensors which are on abnormal skin active, therefore, prolonging the network life time. The size of the aggregation tree grows and shrinks with that of the area and subareas of the affected skin.
	\item An algorithm to estimate the shape of the abnormal area of the skin is proposed for resource limited skin sensors.
	\item The proposed system is demonstrated using various spatio-temporal wound models. 
\end{itemize}

\section{Background \& Related Work}
\label{related}
In this section, a brief overview of skin-on-lab sensing paradigm is presented and then related work on the topic of skin health monitoring is summarized.
\subsection{Skin-On-Lab Paradigm}
Skin-on-lab technology~\cite{doi:10.1021/acsnano.7b04898,doi:10.1002/adma.201504150} involves monitoring and diagnostics sensors that can be attached to the skin. Sensors are flexible and stretchable since they must meet the same characteristics of the skin. Those sensors can be independent standalone sensors or a group of sensors that can be integrated in the skin~\cite{doi:10.1002/adhm.201300220}. Currently, they can be integrated as fabricated temporary tattoos~\cite{doi:10.1002/smll.201402495}, a combination of hard-soft off the shelf electronic components~\cite{Xu70} and functional substrates~\cite{doi:10.1002/smll.201400483}. The sensors can report on many diagnostic signals which depend on the sensing ability and more importantly, the penetration depth of the sensors. Sensors can report on bacterial load, blood analysis, pH level, temperature, glucose level, hydration, oxygen level, bio-potentials, heart rate etc. which can be used to not only measure the health of the skin but also the other body ailments. Clearly, skin-on-lab holds the potential to revolutionize the healthcare with the possibility of integration of drug delivery system on the platform. 

\subsection{A Review of Past research on Skin Health Monitoring}
There have been a plethora of work on health monitoring using BAN; however, skin health monitoring using a body area network has been an overlooked topic perhaps because of lack of skin sensors. A skin-like continuous wound healing monitoring platform is developed chronic wound management which can be laminated at wounds~\cite{doi:10.1002/adhm.201400073}. There have been developments in smart-phone-based applications for skin monitoring~\cite{doi:10.12968/jowc.2016.25.4.177, doi:10.1111/bjd.18152} and cancer detection which can be found in~\cite{CHAO2017551}. There are also skin monitoring techniques for monitoring specific bio-signals from the specific parts of the body, for example,~\cite{Lavery2642} presents a study on home monitoring of foot skin temperatures to detect ulcers for diabetic patients. A smart home system based approach using in-built sensors like motion sensors and cameras is proposed to monitor skin diseases~\cite{Po_ap_2018}, however, such approaches fail to be ubiquitous. 

It is worth noting that the past skin health monitoring approaches neither use body area network nor are adaptive to the dynamic changes in affected area of the skin. This paper utilizes BAN idea to design an energy efficient skin health monitoring system which adapts to the changing skin health.


\section{Skin Monitoring System}
\label{Segalg}

\begin{figure}[htp] 
\centering

\includegraphics[width=0.43\textwidth, height=0.3\textheight]{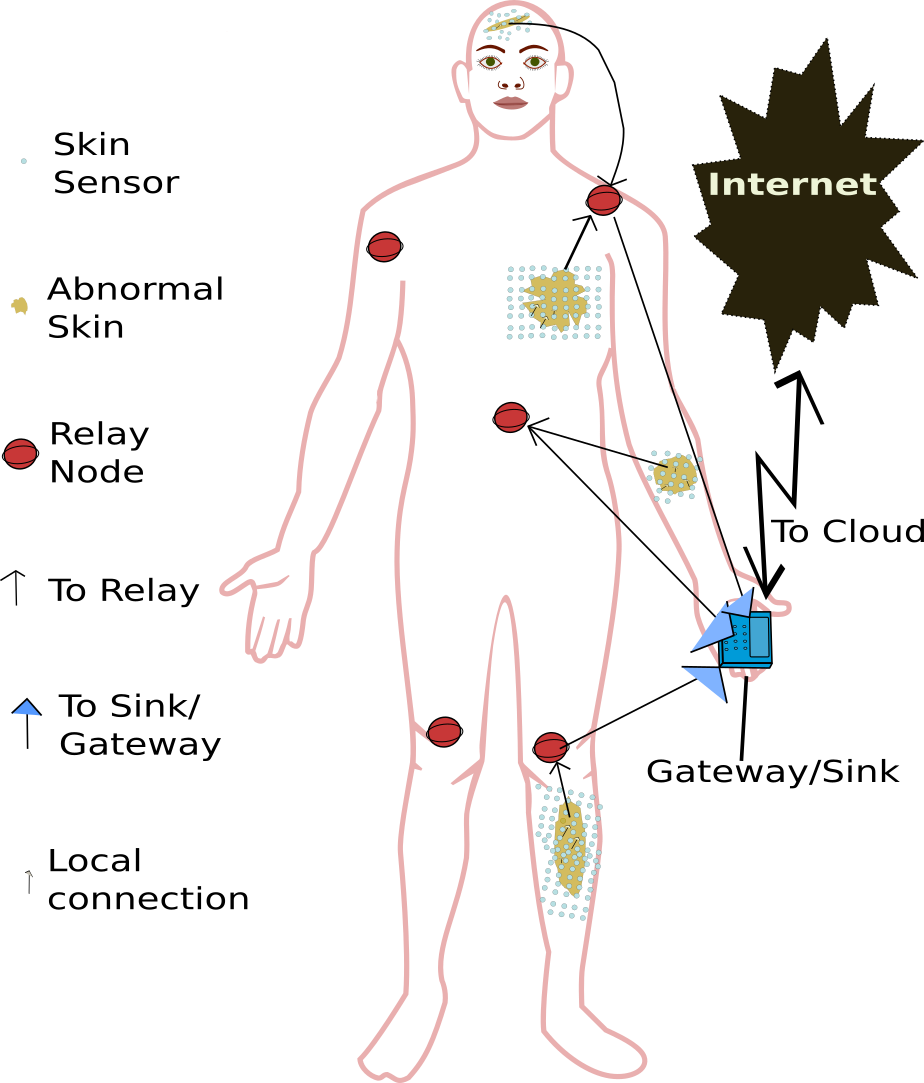} 
\caption{ Body Area Network System architecture}
  \label{BANArch}
\end{figure}\subsection{System Model}
The system architecture is shown in Figure~\ref{BANArch}. Skin sensor nodes are integrated in the skin sufficiently covering the sufficiently large area of skin which requires monitoring. Skin sensors are tiny devices which are independent and capable of limited computation and communication. It is assumed that since these devices are very small, energy consumption is an important concern.  Because of this concern, strategically placed relay nodes are also there to improve the network life time significantly (See~\cite{10.1145/2833312.2849560,6363949}),  although the proposed system can also operate in a multihop fashion and report the data all the way to the sink nodes, however it will be a huge drag on the battery power too. Skin sensors nodes covering the affected area of the skin create spanning trees among themselves for the purpose of data aggregation at the local root node of the tree. The root nodes can pass on the information to relay node directly. Furthermore, relay nodes send the collected information to the sink node which serves as a gateway. The sensed data is sent over the Internet to the remote repository where it is available for further processing and analysis.   

Wireless sensor nodes are integrated into the skin. Sensor nodes are unique and aware of their locations. They are limited by battery power, transmission range and sense the parts of the skin they are integrated into.  Sensors form a network which is modeled as a graph, $G(V,E)$, where $V$ and $E$ are sets of all the sensor nodes and connections, respectively. Sensors can communicate with other body sensors and the relay nodes. 
\subsection{A Distributed Skin Monitoring Algorithm}
There are three different length messages types, $STATUS$, $LOCATION$ and $CHANGE$ which can be distinguished by examining the first couple of bits in the broadcast messages achieving minimal communication energy cost.
\begin{algorithm} [hpt]\small
	\caption{UpdateStatus to neighbours: Code for node $n$}
	\label{algorithmk}
	\begin{algorithmic}[1]
	\State \textbf{Description of symbols:} 
		\State $rID$: root ID
		\State $rE$: root energy level
		\State $n.ID$: Node $n$'s self ID
		\State $n.E$: Node $n$'s self energy level
		\State $n.rID$: Node $n$'s root ID
		\State $n.rE$: Node $n$'s root energy level
		\State $n.pID$: Node $n$'s parent ID
		\\
		\State \textbf{Data Format:} 
\State		$neighboring\_node$: List of neighbors
\\
\State \textbf{Initialize:} n.pID = n.ID, n.rID=n.pID, n.rE = n.E
\\
\State  \textbf{\textit{procedure}} {\textbf{ UpdateStatus():}}
		\For{each $neighboring\_node$}
		\State \textbf{broadcast} \emph{$<$ STATUS, n.ID, rID, rE $>$}
		\EndFor
	\end{algorithmic}
\end{algorithm}


\subsubsection{Energy Efficient Tree Construction}
The nodes run the procedure $UpdateStatus$ as shown in Algorithm~\ref{algorithmk}. Using $UpdateStatus$ function, the sensors sensing the inform the neighbors of its root id and root energy level in the $STATUS$ messages. All active nodes store the root ID and root Energy level of the tree it is a part of. When the algorithm begins, these variables are initialized to its own ID and energy levels.

\begin{algorithm} [hpt]\small
\caption{Energy Efficient Tree: Code for Node $n$}
	\label{algorithml}
	\begin{algorithmic}[1]
	\State \textbf{Description of symbols:} 
		\State $i.as$: Neighbor $i$'s aliveness status
		\State $recharge$: delay constant
		\\
		\State \textbf{Initialize:} for each neighbor $i$, $i.as$=0
		\\
		\State  \textbf{\textit{procedure}} {\textbf{ testNsetP():}}
		\State On receipt of $<$ rID, rE $>$ from $i$:
		\State $i.as$ =$i.as+recharge$
		\If{$i.as$ $>$ $max$}
		\State \emph{i.as = max}
		\EndIf
		\If{$rID$ $=$ $n.ID$}
		\State \emph{setParam(n,n,n)}
		\State Return.
		\EndIf
		\If{$i.rE$ $>$ $n.rE$ OR ($i.rE$ $=$ $n.rE$ AND $i.rID$ $<$ $n.rID$)}
		\State \emph{setParam(n,i,i)}
		\EndIf
		\\
		\State \textbf{\textit{function}} \textbf{setParam($n1, n2$):}
\State \emph{n1.pID = n2.pID}
\State \emph{n1.rE=n2.E}
\State \emph{n1.rID = n2.rID}
	\end{algorithmic}
\end{algorithm}

\begin{algorithm} [hpt]\small
	\caption{Report Border: Code for node $n$}
	\label{algorithmm}
	\begin{algorithmic}[1]
	\State \textbf{Description of symbols:} 
		\State $n.loc$: Node $n$'s location
		\\
	\State  \textbf{\textit{procedure}} {\textbf{ BorderNodeSend():}}
		\State $flag$ = FALSE
		\For{each neighbor $i$}
		\If{$i.as$ $=$ 0}
		\State $flag$ = TRUE
		\Else
		\State $i.as = i.as-1$
		\If{$n.pID$ $=$ $i.ID$ AND $i.as$ $=$ 0}
		\State \emph{setParam(n,n)}
		\State Return.
		\EndIf
		\EndIf
		\EndFor
		\If{flag $=$ TRUE \textbf{AND} $n.pID$ $\neq$ $n.ID$}
		\State \textbf{broadcast} $<LOCATION, n.loc,n.PID >$
		\EndIf
	\end{algorithmic}
\end{algorithm}


\begin{algorithm} [hpt]\small
	\caption{Abnormal Skin Boundary: Code for node $n$}
	\label{algorithmn}
	\begin{algorithmic}[1]
	\State \textbf{Description of symbols and Notations:} 
	\State $b$: $<$Boundary node, Boundary node location$>$
	\State $<d_{new}, d_{old}>$: $<$new distance, old distance$>$
		\State $fDistAngle()$: returns $<$distance, Angle$>$ between two points 
		\State $maxDir$: maximum number of directions allowed
		\State $threshold$: threshold to detect increase or decrease in border
		\State $angle$: The slope of the line joining root node and $b$ node
		\State $angleposition$: Boundary point sampling Angle interval 
			\State $bll$: List at root to store all boundary nodes' information.
			\State $T_{interval}$: Reporting interval to Relay node.
		\\
		\State  \textbf{\textit{procedure}} \textbf{ CollectBoundary\_ReportChange():}
		\If{$n.loc$ $=$ $b.loc$}
		\State Return
		\EndIf
		\\
		\If{$n.pID$ $=$ $n.ID$}
		\State \textbf{broadcast} $<$ LOCATION, $n.pID, n.loc$ $>$
		\Else
		\State $d_{old}$ = $d_{new}$
		\State $<$$dnew$, $angle$$>$ = findDistAngle($b.loc$, $n.loc$)
		\If{$angle$ = $<$ $\pi*DoF$ $>$}
		\State $angle$ =$<$ 0 $>$
		\EndIf
		\State $angleposition$ = \textbf{round}$(angle*MaxDir/(\pi*DoF))$
		\\
		\If{ $b.angle$ exists at $angleposition$}
		\State Set coordinates at $angleposition$ to $<$ $b.loc$ $>$
		\Else
		\If{$d_{new}$ $>$ $threshold$ \( \times \) $d_{old}$}
		\State  \textbf{Report} $<$ CHANGE, $n.loc$ $>$, $b.loc$, + $>$
		\EndIf
		\If{$dnew$ $<$ $threshold$ \( \times \) $dold$}
		\State  \textbf{Report} $<$ CHANGE, $n.loc$ $>$, $b.loc$, - $>$
		\EndIf
		\State  $bll.store$($<$ $b.loc$, $angleposition$ $>$)
		\EndIf
		\\
		
		\State On every $T_{interval}:$\\
		          \ \ \ \ \ \ \ \ \ \ \ \ Send uniform samples of $bll$ to the relay node.
		\EndIf
		
	\end{algorithmic}
\end{algorithm}

The processing of received $STATUS$ messages and construction of tree are shown in Algorithm~\ref{algorithml}. Since microscale sensors are assumed to be independent and resource constrained, the energy consideration is extremely important. The algorithm is designed to consume minimal storage space. Nodes are assumed to be alive for the next certain number of rounds if the message is received in the current round. This information is maintained in alive status variable - $i.as$ which is maintained by every node for every neighbor nodes.  Once the $send$ messages (sent in Algorithm~\ref{algorithmk}) is received from the neighbor, alive status is extended by $recharge$ duration. The maximum value of $i.as$ is $max$ (see Algorithm~\ref{algorithml}) to stop overflow error as well as limit the assume duration of liveness of neighboring nodes. If the root id of the sender is the id of the receiver, set the parent id, root id and root energy of the receiver to its own values ($n.pd$ is the id of the parent) to the receiver (line 21). If the root energy of the sender is greater than the root energy of the receiver or the energies are the same and the root id of the sender is less than the root energy of the receiver, set the parent id to the received sender id, and corresponding root energy and root id of the receiver to those sent by the sender. Therefore, in the case of equal energy, the symmetry is broken using the node id. 

The proposed algorithm does not require tracking of children. Sent messages are directed to the parent. The parent identifies that the message is meant for it by observing its own id in the received messages. This helps us minimize the storage memory requirement in the case of resource constraint devices and dense sensor deployment.

\subsubsection{Estimation of Size and Shape of Abnormal Skin} 

Reporting the shape and size of and dynamic changes in the abnormal skin area are done using the Algorithm~\ref{algorithmm} and~\ref{algorithmn}. The Algorithm~\ref{algorithmm} describes the process of reporting the boundary. Only nodes forming the boundary of abnormal skin area report the location data, other nodes, still monitoring but do not report. The criteria for a node to be a boundary node is that at least one of its neighbor is asleep/inactive. A sleep node indicates that there is no abnormal condition there, therefore, the active node which is the neighbor of the inactive node is a border node. This is implemented by checking the alive status (line 7), if the alive status is 0 then node is inactive and the current node $n$ is the boundary node. In that case,  given node is not the root node then $n$ broadcasts the $LOCATION$ message (line 18). Algorithm~\ref{algorithmm} ensures minimal communication making it an energy efficient scheme.  

The Algorithm~\ref{algorithmn} depicts the reporting on abnormal skin boundary as well as changes in boundaries as compared to previous reporting. All boundaries are collected and stored in list $bll$ at the root. Then the boundary nodes are sampled at an angle interval, given by $angle$ and total interval is given by $MaxDir$. $angle$ is computed by dividing the maximum angle by $\pi \times DoF$ where $DoF$ is degree of freedom by $MaxDir$ as seen in line  results in the closest approximation to $angle$. This number is used to approximate the growth or shrinkage of the wound along a given direction. For example, if we have 20 pairs of (x,y) coordinates to store the results, we divide the angle by 18 (=360/20) and round the answer to the nearest integer to find the coordinate position to store the boundary. When the root reports a new boundary node, new distance is calculated from the root to the old boundary. If the change in distance is above or below a certain threshold, then if the new distance is greater, \textbf{Report} change in that direction, \emph{+} for growth and \emph{-} for shrinkage (lines 30 - 35). Periodic updates are done by storing the boundary information in $bll$ (line 36), storage for boundary information and sending them at regular intervals (lines 39-40).

%
%
%

\subsection{Complexity Analysis} 
\subsubsection{Time Complexity}
Every node must follow the highest energy node as its root, therefore, they must receive this information in a multiple fashion. Time complexity depends on the distance between the highest energy nodes and the farthest nodes in number of hops. Since highest distance between any pair of nodes is the diameter of the monitoring area, therefore, time complexity of Energy Efficient Tree algorithm (Algorithm~\ref{algorithml}) is $O(D_{highest})$ where $D_{highest})$ is the largest diameter among all the sub networks deployed in the monitoring region. Likewise, the time complexity of the Algorithm~\ref{algorithmm} is also $O(D)$ where $D$ is the diameter of the monitoring subarea.
\subsubsection{Message Complexity}
The nodes placed on normal skin never send any message. There are a total of N spanning tree formation in progress and every formation requires at most O(D) messages traversing through a network with diameter $D$. Therefore, the message complexity of Tree formation algorithm (Algorithm~\ref{algorithml}) is $O(ND_{network})$ where $D_{network}$ is the summation of all the diameters of the subnetworks formed on the monitoring area. Furthermore, in the reporting phase Algorithm~\ref{algorithmm}, the number of messages originated depends on the $p$, the perimeter of the monitoring region and the messages must reach the root node, therefore, the message complexity in the reporting phase is $O(Dp)$ where $D$ is the diameter of the abnormal skin area. 

\section{Result and Discussion}
\label{Evaluation}

\subsection{Simulation Setup}

\begin{figure*}[h] 
  \centering
  \subfigure[GunShot: t=0]{\label{GunShott0}
   \includegraphics[width=0.235\textwidth]{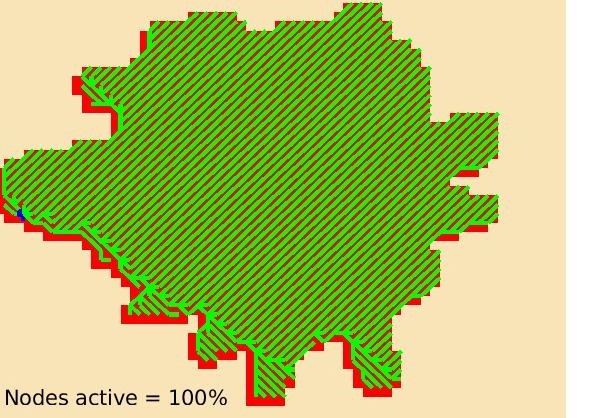}} 
  \subfigure[GunShot: t=1]{\label{GunShott1}
   \includegraphics[width=0.235\textwidth]{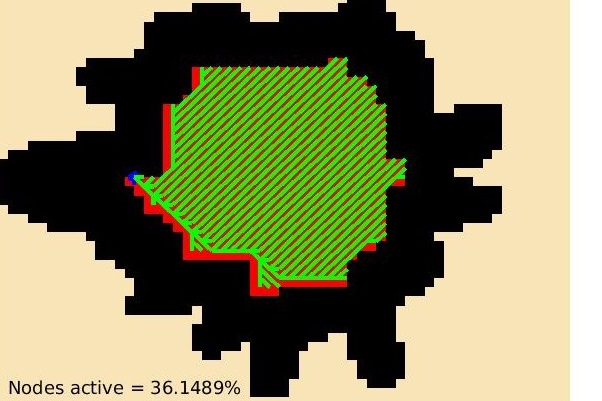}} 
	\subfigure[GunShot: t=2]{\label{GunShott2}
   \includegraphics[width=0.235\textwidth]{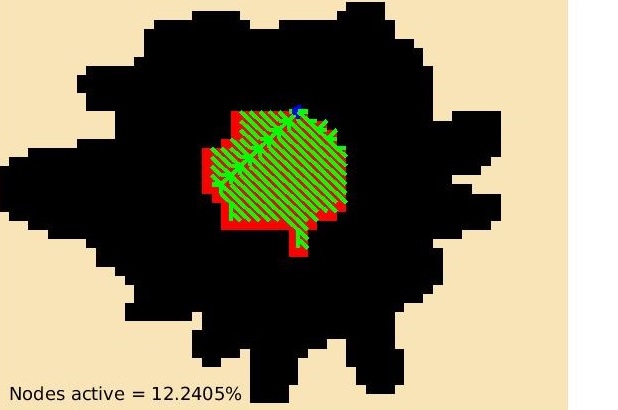}} 
   \subfigure[GunShot: t=3]{\label{GunShott3}
   \includegraphics[width=0.235\textwidth]{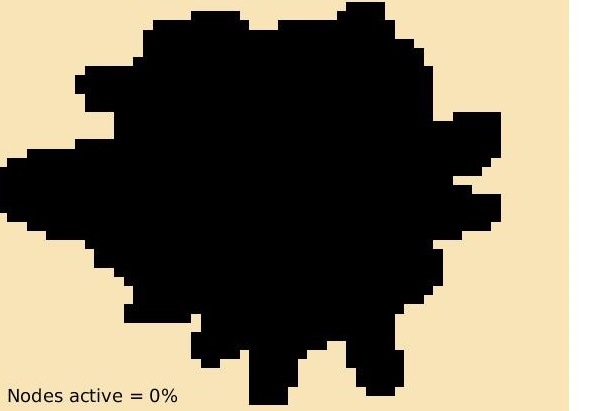}\par\medskip} 
	\subfigure[Scratch: t=0]{\label{Scratcht0}
	   \includegraphics[width=0.235\textwidth]{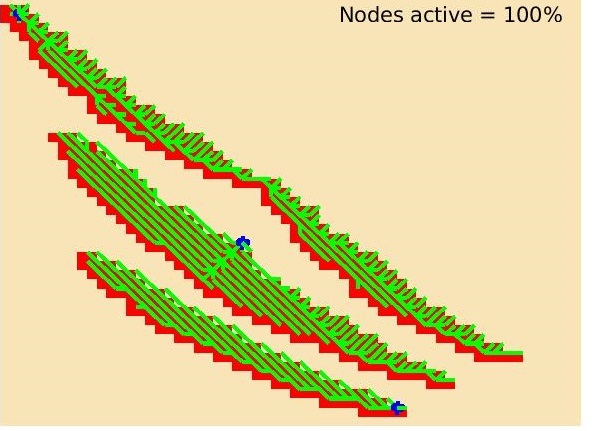}} 
  \subfigure[Scratch: t=0]{\label{Scratcht1}
   \includegraphics[width=0.235\textwidth]{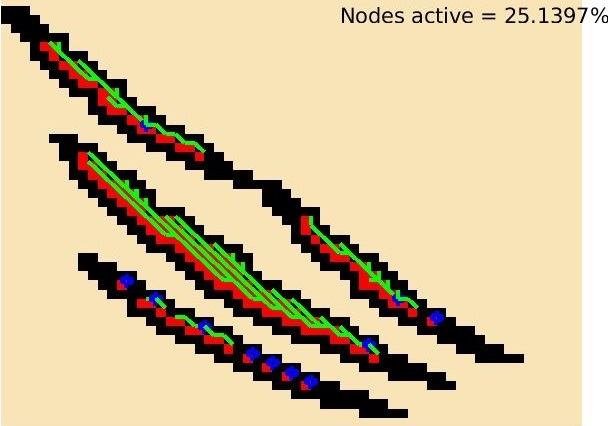}} 
	\subfigure[Scratch: t=2]{\label{Scratcht2}
   \includegraphics[width=0.235\textwidth]{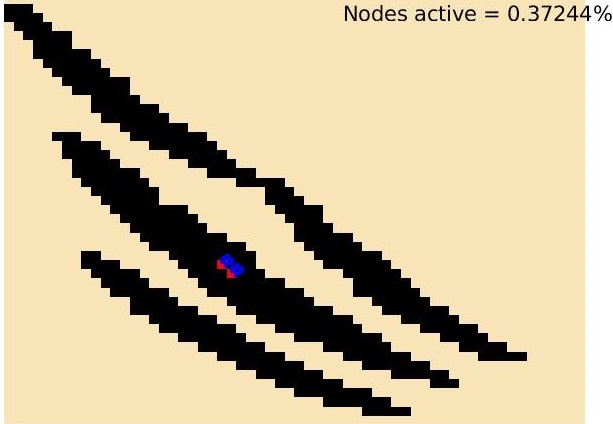}} 
   \subfigure[Scratcht3]{\label{Scratcht3}
   \includegraphics[width=0.235\textwidth]{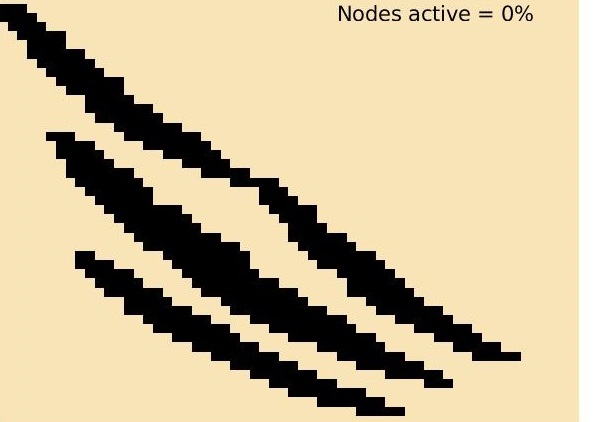}\par\medskip} 
	
	\subfigure[Oval Shape: t=0]{\label{OvalShapet0}
   \includegraphics[width=0.235\textwidth]{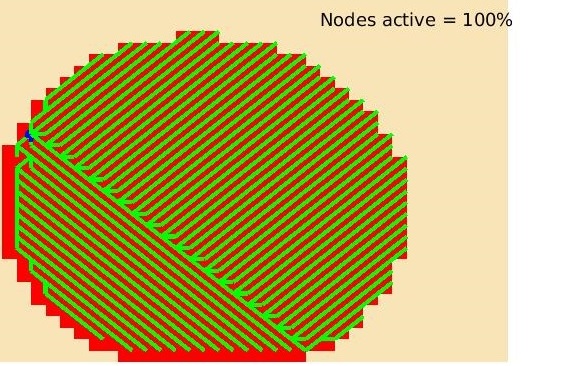}} 
  \subfigure[Oval Shape: t=2]{\label{OvalShapet1}
   \includegraphics[width=0.235\textwidth]{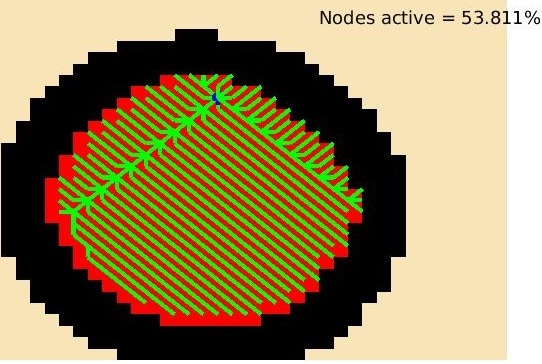}} 
	\subfigure[Oval Shape: t=2]{\label{OvalShapet2}
   \includegraphics[width=0.235\textwidth]{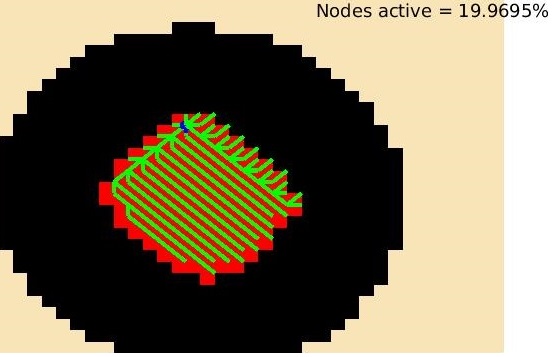}} 
   \subfigure[Oval Shape: t=3]{\label{OvalShapet3}
   \includegraphics[width=0.235\textwidth]{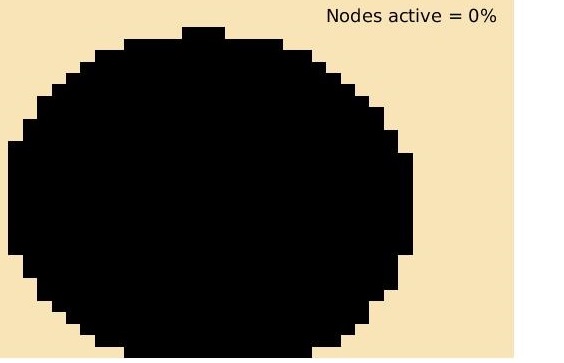}\par\medskip} 
   \caption{Energy efficient tree covering the wound with edges shown in green color at four diffent time snapt shots. Larger green dots and blue dots are parent and root nodes respectively. Active and healed wound portions are shown in red and black.}
  \label{asynchange}
\end{figure*}


The simulation uses the energy model designed after nRF24L01  2.4GHz transceivers for lossy intrabody environment from the implant to the on body as described in~\cite{Liao_2018}. The parameters are shown in table~\ref{parameters}. For demonstration purpose, 2D co-ordinate system is used. A set of skin wounds is chosen to test the algorithm because wounds exhibit suitable behavior of shrinkage or spread. We utilize a simple wound healing model as described in~\cite{woundmodel1990}. Two different wound types which are prevalent in our world are chosen. One is induced by gun shots and the other is caused by cut or scratch by sharp objects as shown in Figure~\ref{asynchange}. The gunshot wound is first let grown and then it shrinks where as wound caused by cut  shrinks rapidly. In addition to these, we also present the results on oval shaped wound which is assumed to be healing smoothly and uniformly. The simulation is conducted using Matlab.
\begin{center} 
\begin{table}[t]
\centering
\caption{Parameter Setup}
\label{setup}
	\begin{tabular}{|c|p{1cm}|p{1cm}|p{1cm}|p{1cm}|p{1cm}|p{1cm}|}
		\hline
		  $E_{trx}$ & $E_{rec}$ & $\epsilon_{amp}$& MaxP & recharge &threshold \\
		\hline		
		16.7nJ/bit & 36.1nJ/bit & 1.97nJ/biy & 20 & 3 &10\%\\

		\hline
	\end{tabular}
	\label{parameters}
\end{table}
\end{center}
\subsection{Demonstration}
\subsubsection{Wound Monitoring with Energy Efficient Tree}
Figure~\ref{asynchange} demonstrates the working of the skin health monitoring algorithm on three different kinds of wounds, caused by gun shot, scratch and an oval shape. Active wounds are colored red whereas healed portion is colored black. The sensors that are not on active would are put to sleep and not shown in the figure. Active wounds are shown in red whereas the healed portion is shown in black. The status of wound and BAN using skin sensors are shown at four different time snapshots. Times t=0, t=1, t=2 and t=3 indicate initial, intermediate, almost healed and completely healed states of wounds, respectively and the corresponding states of skin-sensor network. 

Active sensors on the wound form a network and shown using green color. The parent nodes and root nodes are shown using thicker green and blue points, respectively. Figures~\ref{asynchange}(a) - (d) show skin sensors deployed on the gun wound. As the wound is healing and shrienking in size, the sensor network is adapting to the the changing wound area. Comparison among Figures{asynchange}(a)-(c) shows that the aggregation tree is also changing which can be observed by different nodes taking on the roles of parent and root nodes. Finally, the sensor nodes turn inactive when the wound is healed completely (Figure~\ref{asynchange}(d)). The scratch wound is shown in Figures~\ref{asynchange}(e)-(h). Scratch would consists of three localized wounds. Accordingly, sking sensors create three localized networks covering localized wounds. Three networks with their own root nodes are formed as nodes are healing, the network is also changing. Figure~\ref{asynchange}(g) shows that a small portion of the middle wound is active and monitored by very few nodes whereas other wounds are completely healed leading to sensors in those regions becoming inactive. Likewise, oval shape wound (Figure~\ref{asynchange}(i)-(l)) shows the similar behavior.    

\subsubsection{Network Life time}
\begin{figure}[h] 
\centering
  \subfigure[Energy Consumption]{\label{EnergyCons}
   \includegraphics[width=0.234\textwidth]{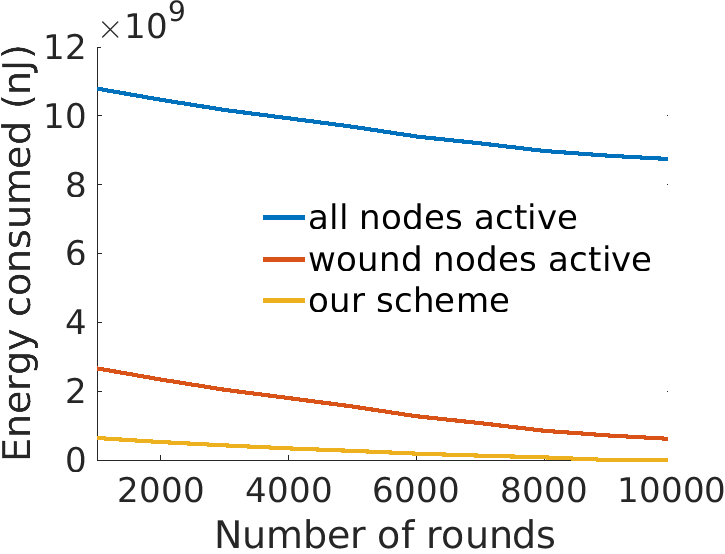}} 
  \subfigure[Dead Nodes]{\label{DeadNodes}
   \includegraphics[width=0.234\textwidth]{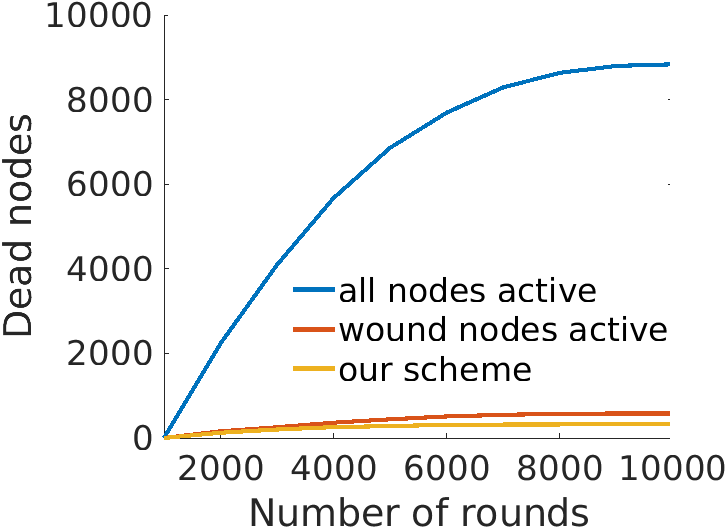}} 
\caption{Energy Usage \& Dead Nodes vs Number of Rounds}
  \label{NetworkLifeTime}
\end{figure}

Network life time is presented in the form of Energy consumption and number of dead nodes in Figure~\ref{NetworkLifeTime}. The proposed scheme is compared with the cases, when all the nodes are active, therefore there is no self adaptation, and when only nodes on the active wound are active, however, there is no adaptation to the changing energy of the nodes. When all the nodes are active all the time, energy consumption is high and there are more dead nodes. Improvement in energy consumption and alive nodes observed when would only nodes are allowed to be active. The proposed scheme with self adaptation mechanism and energy efficient tree significantly outperforms the other schemes which exhibits lesser energy consumption and lower number of dead nodes, effectively, prolonging the network lifetime. 


\section{Conclusion \& Future Work}
\label{conclusion}
The paper presents the first distributed skin health monitoring system using WBAN for lab-on-skin sensing paradigm. The algorithm adapts to the spread and/or shrinkage of the affected area dynamically and also to the new areas of skin abnormalities. An energy efficient aggregation tree topology is created for data collection from the monitored area. The root of the aggregation tree is the highest energy level sensor node which is selected dynamically extending the network life time. The algorithm is demonstrated using a variety of wound profiles. The proposed system promotes telemedicine long term data collection that has a potential to promote data driven health care research and development.

A large scale study of the proposed system exploring its performance over a diverse set of skin conditions is left for future work. Also, further study should explore the impact of lossy environment on the performance. The adaptation of the proposed monitoring system in the presence of strategically placed energy harvesters is worth exploring in the future. Furthermore, the integration of drug delivery system and proposed system is very valuable which can automate the diagnosis and treatment methods.

%

\bibliographystyle{IEEEtranS}
\bibliography{reference,bibfile_Kumar}

\end{document}